\documentclass[jpa]{iopart}
\usepackage{amsfonts}
\usepackage{amssymb}

\usepackage[dvips]{graphicx}
\usepackage{dcolumn}
\usepackage{bm}
\usepackage{psfrag}

\begin{document}

\title[Detecting Generalized Synchronization by A Kernel-based Approach]{Detecting Generalized Synchronization Between Chaotic Signals: A Kernel-based Approach}   
\author{Hiromichi Suetani$^{1, 3}$, Yukito Iba$^{2}$ and Kazuyuki Aihara$^{3, 1}$}
\ead{$^{1}$suetani@aihara.jst.go.jp}
\address{$^{1}$Aihara Complexity Modelling Project, ERATO, JST, 
3-23-5  
Uehara, Shibuya-ku, Tokyo 151-0064, Japan}
\address{$^{2}$The Institute of Statistical Mathematics, 
4-6-7  
Minami-Azabu, Minato-ku, Tokyo 106-8569, Japan}
\address{$^{3}$Institute of Industrial Science, The University of Tokyo, 
4-6-1  
Komaba, Meguro-ku, Tokyo 153-8505, Japan}

\begin{abstract}
 A unified framework for analyzing generalized synchronization in coupled chaotic systems from data is proposed.
The key of the proposed approach is the use of the {\it kernel methods} recently developed in the field of machine learning. 
Several successful applications are presented, which show the capability of the kernel-based approach for detecting generalized synchronization, and dynamical change of the coupling strength between two chaotic systems can be captured by the proposed approach. 
It is also discussed how the kernel parameter is suitably chosen from data.      
\end{abstract}
\submitto{\JPA}
\maketitle 

\section{Introduction}
Synchronization of chaotic systems has been an active research area in recent years~\cite{Piko01}.
Now the notion of synchronization of chaotic systems is extended far beyond complete synchronization between identical systems\cite{Fuji83,Piko84,Peco90},  and various kinds of nonlinear synchronization have been proposed~\cite{Piko01}.
%
%
A significant extension is {\it generalized synchronization} (GS)~\cite{Rulk95}, which is defined by a time-independent nonlinear functional relation $y = \Psi (x)$ between the states $x$ and $y$ of two 
systems $X$ and $Y$.
%

Experimental detection and characterization of GS from observed data is a challenging problem, especially in biology; e.g., for study on nonlinear interdependence observed in binding of different features in cognitive process~\cite{Farm98} and epilepsies in the brain~\cite{Quye99,Arnh99}.

In unidirectionally coupled systems,  
a way to detect GS is to make an identical copy $Y'$ of the response system $Y$ driven by the common signal from the driver system $X$,  then investigate whether  orbits of both $Y$ and $Y'$ coincide after transient~\cite{Koca96,Abar96}.
However, it is very difficult to prepare an identical copy of the response system, and  this approach does not give any information on  the structure of the synchronization manifold ${\cal M} = \{(x, y) : y=\Psi (x)\}$ in the state space.  

On the other hand, several indices  have been proposed~\cite{Rulk95,Quye99,Arnh99,Schi96,Quia00,Stam02},  to quantify  nonlinear dependence between $X$ and $Y$ based on a local relation between them.  
While the usefulness of these indices were shown in some examples~\cite{Quye99,Arnh99,Schi96}, those approaches are still ad-hoc,   lacking systematic methods for unifying the local relation of different regions in the state space.  

Recently, {\it kernel methods} have been attracting much attention in the machine learning community as powerful tools for analyzing data with high nonlinearity ~\cite{Scho98,Mull01,Scho02,Shaw04}. 
In this paper, we propose a novel approach for analyzing GS based on {\it Kernel Canonical Correlation Analysis} (Kernel CCA)~\cite{Lai00,Akah01,Melz01,Bach02} which is a version of  the kernel methods~\cite{Sue06}\footnote{Preliminary results are reported in a conference proceedings~\cite{Sue06} with a simpler example of detection of GS. Here, we give comprehensive treatments with examples of practical interests
}.
We will demonstrate that the Kernel CCA provides a suitable index for measuring the nonlinear interdependence, and gives global coordinates characterizing the synchronization manifold  ${\cal M}$ of GS.
Note that the latter  has not been addressed in conventional studies on the analysis of GS.  
The proposed method does not require explicit knowledge on the underlying equations behind time series.
Although we restrict our attention to the analysis of numerical experiments here, it is also applicable to experimental data.
 
The present paper is organized as follows.
In section~2, we introduce a formulation of the Kernel CCA.
Sections~3 and  4 provide several successful applications of the Kernel CCA for treating GS.
In section~5, it is demonstrated that the nonstationary change of the coupling strength between two chaotic systems can be captured by the proposed approach. 
In section~6, we discuss an issue on the choice of kernel parameters. 
Our conclusion is given in section~7.  

\section{Kernel Canonical Correlation Analysis}
Canonical Correlation Analysis (CCA)
is a multivariate analysis method to find a pair of vectors that maximize the correlation coefficient between projections of signals onto them~\cite{Hote36}.
CCA is useful for detecting a linear relation between a pair of  multidimensional data sets, but cannot deal with the case like GS where there is a high nonlinear relation between two data sets. 
Extending the ability of CCA for analyzing data with nonlinearity, kernelization of CCA has been proposed by several researchers~\cite{Lai00,Akah01,Melz01,Bach02}.
An intuitive formulation of the Kernel CCA is as follows.
For a pair of multivariate variables $(x, y)$ with $x\in {\mathbb R}^p$ and $y \in {\mathbb R}^q$, the Kernel CCA seeks a pair of nonlinear scalar functions  $f: {\mathbb R}^p \to {\mathbb R}$ and $g: {\mathbb R}^q \to {\mathbb R}$ such that an estimator of the correlation coefficient 
\begin{eqnarray}
\label{rho_F}
\rho_{\cal F} = 
\frac{{\sf cov}(f(x), g(y))}
{\sqrt{{\sf var}(f(x))}\sqrt{{\sf var}(g(y))}}
\end{eqnarray}
between $f(x)$ and $g(y)$ is maximized.

Methods based
on the expression~(\ref{rho_F}) of the generalized
correlation coefficient have a long history~\cite{Buja90,Brei85}.
However the use of the kernel approach
gives a notable progress in the subject, with
a remarkably simple and flexible way of treating~(\ref{rho_F}). 

The essence of the kernel methods is an assumption that nonlinear functions $f$ and $g$ are well approximated by linear combinations of {\it kernels} on  data points $(x_n, y_n)$, 
\begin{eqnarray}
\label{fg}
 f(x) = \sum_{n=1}^N \alpha_n k(x_n, x), \quad g(y) = \sum_{n=1}^N \beta_n k(y_n, y), 
\end{eqnarray}
where $\{(x_n, y_n)\}_{n=1}^N$ denotes  a training data set of $(x, y)$. 

Examples of kernels are 
$k(x, x') = {\rm e}^{\frac{-\Vert x - x'\Vert^2}{2\sigma^2}}  \ ({\rm Gaussian})$, 
$((x\cdot x')+\theta)^d ) \ ({\rm Polynomial})$,  and $\tanh (\eta (x\cdot x')+\theta)  \ ({\rm Sigmoidal})$~\cite{Mull01,Scho02,Shaw04}. 
A rich representation of nonlinear functions is allowed 
by suitably chosen weight coefficients $\{ \alpha_n\}_{n=1}^N$, $\{ \beta_n\}_{n=1}^N$ and kernel parameters. 
At first sight, the dependence of the expressions for $f$ and $g$ on the given data $\{(x_n, y_n)\}_{n=1}^N$  brings some ad-hoc nature to the method. 
The assumption is, however,  equivalent to the assumption that $f$ and $g$ are minimum norm functions in a Reproducing Kernel Hilbert Space (RKHS).
 It is proved by invoking {\it representer theorem} in the theory of RKHS~\cite{Mull01,Scho02,Shaw04}.

Substituting  (\ref{fg}) into  (\ref{rho_F}),  and replacing covariance ${\sf cov} (\cdot, \cdot)$ and variances ${\sf var} (\cdot)$ with empirical averages over the data set $\{(x_n, y_n)\}_{n=1}^N$,  the Kernel CCA reduces to the maximization of 
\begin{eqnarray}
\label{rho_2}
\rho_{\cal F} =
\frac{{}^t\alpha K_X K_Y \beta}{\sqrt{{}^t\alpha K_X^2\alpha}\sqrt{{}^t\beta K_Y^2 \beta}}, 
\end{eqnarray}
where $\alpha={}^t (\alpha_1, \alpha_2, ..., \alpha_N)$, $\beta={}^t (\beta_1, \beta_2, ..., \beta_N)$, and $K_X, K_Y$ are the {\it Gram matrices} $(K_X)_{i, j} = k(x_i, x_j)$ and $(K_Y)_{i, j} = k(y_i, y_j)$ defined with the given sample.
The Gram matrix contains relevant topological information of data points in the state space.
For simplicity,
we tentatively assume that the averages of $\{f(x_n)\}_{n=1}^N$ and $\{g(y_n)\}_{n=1}^N$ are zero.
Later we will show that subtraction of the averages from data can be done in an implicit  way and  does not cause any technical difficulty.

A naive generalization of CCA leads to the maximization of ${}^t \alpha K_XK_Y\beta$ subject to  ${}^t\alpha K_X^2 \alpha = {}^t\beta K_Y^2 \beta = 1$. 
The corresponding Lagrangian is
\begin{eqnarray}
\label{Lagrange_0}
{\cal L}_0 = {}^t \alpha K_XK_Y \beta - \frac{\rho_X}{2} ({}^t\alpha K_X^2 \alpha -1) 
- \frac{\rho_Y}{2} ({}^t\beta K_Y^2 \beta -1), 
\end{eqnarray}
where $\rho_X$ and $\rho_Y$ are Lagrange multipliers. 
Taking derivatives of ${\cal L}_0$ with respect to $\alpha$ and $\beta$, and $\rho_X = \rho_Y = \rho$ from the constraint, the Kernel CCA is formulated as the following generalized eigenvalue problem: 
\begin{eqnarray}
\label{KCCA_1}
 \left (
\begin{array}{cc}
0 & K_X K_Y \\
 K_Y K_X & 0
\end{array}
\right )
\left(
\begin{array}{c}
\alpha \\
\beta
\end{array}
\right)
=
\rho
\left (
\begin{array}{cc}
K_X^2 & 0 \\
 0 & K_Y^2
\end{array}
\right )
\left(
\begin{array}{c}
\alpha \\
\beta
\end{array}
\right).
\end{eqnarray}
There is, however,  a difficulty when the Gram matrices $K_X$ and $K_Y$ are invertible. 
By multiplying both sides of  (\ref{KCCA_1}) by $(K_X^{-1}K_X^{-1}, K_Y^{-1}K_Y^{-1})$ from the left hand side, we have
\begin{eqnarray}
\label{invert_case_alpha}
\alpha=\frac{1}{\rho}K_X^{-1}K_Y\beta, \\
\label{invert_case_beta}
\beta=\frac{1}{\rho}K_Y^{-1}K_X\alpha.
\end{eqnarray}
Substituting  (\ref{invert_case_beta}) into   (\ref{invert_case_alpha}), we obtain
\begin{eqnarray}
\label{identity}
I\alpha = \rho^2\alpha, 
\end{eqnarray}
which gives a trivial solution $\rho=\pm 1$. 
Such a situation is not uncommon, because the Gram matrix is invertible when a Gaussian kernel is used and data points are distinct each other.
Thus the naive kernelization (\ref{KCCA_1}) of CCA  does not provide useful information on the nonlinear dependence. 

To overcome this problem we introduce small regularization terms in the denominator of the right hand side of  (\ref{rho_2}) as 
\begin{eqnarray}
\label{rho_3}
\rho_{\cal F} =
\frac{{}^t\alpha K_X K_Y \beta}{\sqrt{{}^t\alpha K_X^2\alpha + \kappa \Vert f \Vert^2_{\cal F}}\sqrt{{}^t\beta K_Y^2 \beta + \kappa \Vert g \Vert^2_{\cal F}}}, 
\end{eqnarray}
where $\Vert f \Vert^2_{\cal F}$ and $\Vert g \Vert^2_{\cal F}$ are quadratic norms of $f$ and $g$ on the RKHS defined as 
 \begin{eqnarray}
\label{norm_on_RKHS}
\Vert f \Vert^2_{\cal F} &=& \sum_{n'=1}^N \alpha_{n'} f(x_{n'})= \sum_{n'=1}^N \sum_{n=1}^N \alpha_{n'} \alpha_n k (x_n, x_{n'})\nonumber \\  &=& {}^t \alpha K_X\alpha, \\
\Vert g \Vert^2_{\cal F} &=& \sum_{n'=1}^N \beta_{n'} g(y_{n'})=\sum_{n'=1}^N \sum_{n=1}^N \beta_{n'} \beta_n k (y_n, y_{n'})\nonumber \\  &=&{}^t \beta K_Y\beta,
\end{eqnarray}
and $\kappa$ is its parameter. 
Note that although the parameter $\kappa\neq 0$ is required for a nontrivial result, the precise value of $\kappa$ is not important\footnote{Recently, statistical consistency of the Kernel CCA, i.e., convergence of the estimators of $f$ and $g$ to the optimal ones is proved when $f$ and $g$ belong to a RKHS, under the condition that $\kappa /N \sim N^{-1/3 + \delta} (0<\delta <1/3)$ with $N\to \infty$. Fukumizu K, Bach F R, and Gretton A 2005 Consistency of kernel canonical correlation analysis, {\it ISM Research Memorandum} No.942}
This insensitivity to the value of $\kappa$  will be confirmed  by a numerical experiment in the next section. 
As well as  (\ref{KCCA_1}),  the maximization of the numerator ${}^t\alpha K_XK_Y\beta$ in  (\ref{rho_3}) subject to ${}^t\alpha K_X^2 \alpha + \kappa {}^t \alpha K_X\alpha = {}^t\beta K_Y^2 \beta + \kappa {}^t \beta K_Y\beta = 1$ reduces to the following generalized eigenvalue problem 
\begin{eqnarray}
\label{KCCA_2}
&&
 \left (
\begin{array}{cc}
0 & K_X K_Y \\
 K_Y K_X & 0
\end{array}
\right )
\left(
\begin{array}{c}
\alpha \\
\beta
\end{array}
\right)
\nonumber 
\\
&& =
\rho
\left (
\begin{array}{cc}
K_X(K_X + \kappa I) & 0 \\
 0 & K_Y(K_Y + \kappa I)
\end{array}
\right )
\left(
\begin{array}{c}
\alpha \\
\beta
\end{array}
\right).
\end{eqnarray}
The first eigenvalue of  (\ref{KCCA_2}) gives the maximal value $\rho_{\cal F}^{\max}$ of $\rho_{\cal F}$ in  (\ref{rho_3}).
$\rho_{\cal F}^{\max}$ is called the {\it canonical correlation coefficient}, and the variables $u=f(x)$ and $v=g(y)$ transformed by $f$ and $g$ are called the {\it canonical variates} of the Kernel CCA.  

So far, the averages of $\{f(x_n)\}_{n=1}^N$ and $\{g(y_n)\}_{n=1}^N$ are 
set to zero.
In fact, it is not necessary to subtract the averages explicitly,
because the subtraction of the means is equivalent to the replacement of the Gram matrices $K$  with
\begin{eqnarray}
\label{Centering}
\tilde{K} = K - 
\frac{1}{N} 
({\bf j} \> {}^t{\bf j}) K - 
\frac{1}{N} 
 K ({\bf j} \>  {}^t{\bf j})  + 
\frac{1}{N^2} 
({\bf j} \>  {}^t{\bf j}) K ({\bf j} \> {}^t{\bf j}),
\end{eqnarray}
where ${\bf j}$ is a $N$-dimensional vector such that each component
equals to the unity~\cite{Mull01,Scho02,Shaw04}.  

In the examples of this paper, the data $\{x_n\}_{n=1}^N$ and $\{y_n\}_{n=1}^N$ are normalized within an unit interval $[0,1]$ before applying the Kernel CCA.

\section{Detecting Generalized Synchronization by Kernel CCA}
The kernel methods such as the Kernel CCA have been mainly applied to problems of pattern recognition~\cite{Shaw04} and bioinformatics~\cite{Scho04}. 
In this paper, we propose that the Kernel CCA can also be a powerful tool for analyzing nonlinear dynamics. 

Voss {\it et al.}~\cite{Voss97} proposed a method
for analyzing dynamical system
by the maximization the expression~(\ref{rho_F}).
It is based on the {\it Alternating Conditional Expectation} (ACE)
algorithm~\cite{Brei85}, where the transformations 
$f$ and $g$ in~(\ref{rho_F}) are restricted to 
a linear combination of univariate functions such as
$
f(x_1,x_2,\ldots,x_p)= \sum_i \Phi_i(x_i) \, {}_,
$
while the Kernel CCA 
allows any nonlinear function $f(x_1,x_2,\ldots,x_p)$ 
of $p$ variables in a RKHS.
This difference makes the proposed method much more
powerful than the one based on ACE, especially
in the cases where nonlinear dependence between 
variables $x_1,x_2, \ldots$ is essential.
Unlike ACE, the Kernel CCA does not need any 
iterative procedure and it is enough to solve a generalized eigenvalue
problem by standard numerical algebra, only once for each set
of data and parameters. 

%
In this section, we study two unidirectionally 
coupled H\'{e}non maps~\cite{Schi96} as an illustrative
example and demonstrate the ability of the proposed approach. \\ 

\subsection{Quantitative Characterization of Generalized Synchronization}
Two unidirectionally coupled H\'{e}non maps are described by the following difference equations: 
\begin{eqnarray}
\label{Henon_1}
&& X:\left \{
\begin{array}{ll}
x_1(t+1) = 1.4- x_1 (t)^2 + 0.3 x_2 (t)  \\
x_2 (t+1) = x_1(t),  
\end{array}
\right. 
\\
\label{Henon_2}
&& Y:\left \{
\begin{array}{ll}
y_1 (t+1) = 1.4- \{\gamma x_1 (t) y_1(t) + (1 - \gamma) y_1(t)^2\} + 0.1 y_2 (t)  \\
y_2 (t+1) = y_1 (t),  
\end{array}
\right.
\end{eqnarray}
where $x(t)=(x_1(t), x_2(t))$ and $y(t)=(y_1(t), y_2(t))$ are state variables at time $t$, and  $\gamma$ denotes the coupling strength between two systems $X$ and $Y$. 

\begin{figure}[hptb]
\begin{center}
\includegraphics[width=12cm]{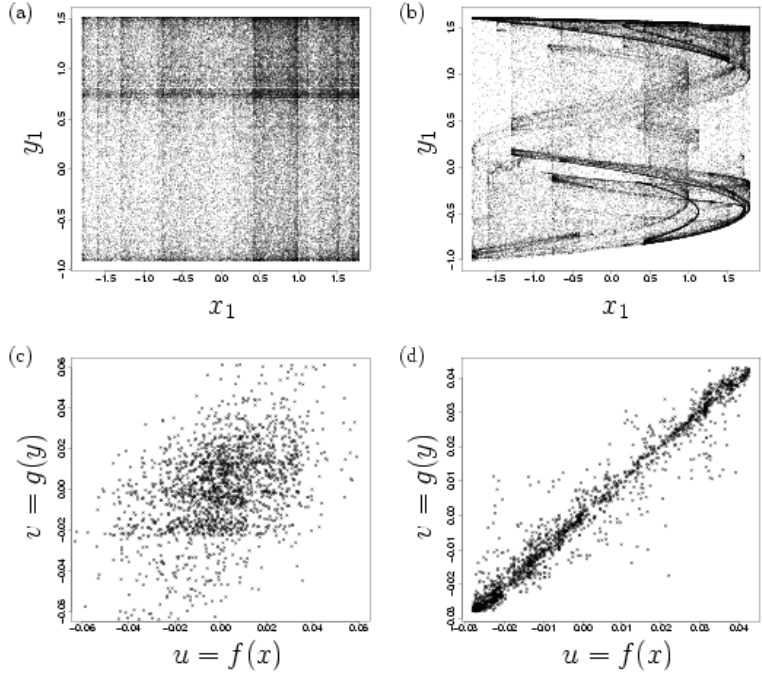}
\caption{(a, b) Projections of strange attractors of the coupled H\'{e}non maps onto the $(x_1, y_1)$ plane with $\gamma=0.0$ in (a) and $\gamma=0.25$ in (b).  For both of (a) and (b), an orbit with length $5\times 10^4$ is used for plotting. (c, d) Scatter plots of the canonical variates of the Kernel CCA with $\gamma=0.0$ in (c) and $\gamma = 0.25$ in (d).}
\label{revise_figure1}
\end{center}
\end{figure}

Figures~\ref{revise_figure1} (a) and (b) show the projections of the strange attractors of  (\ref{Henon_1}) and (\ref{Henon_2}) with $\gamma = 0.0$ and $0.25$ onto the $(x_1, y_1)$ plane, respectively.
When the coupling strength $\gamma$ is large, a complicated driver-response relation with high  nonlinearity is formed between two different chaotic dynamics.   
It can be identified in figure~\ref{revise_figure1} (b) with a visual inspection, but it is not easy to represent it by naive statistical  tools such as correlation coefficients. 

\begin{figure}[hptb]
\includegraphics[width=12cm]{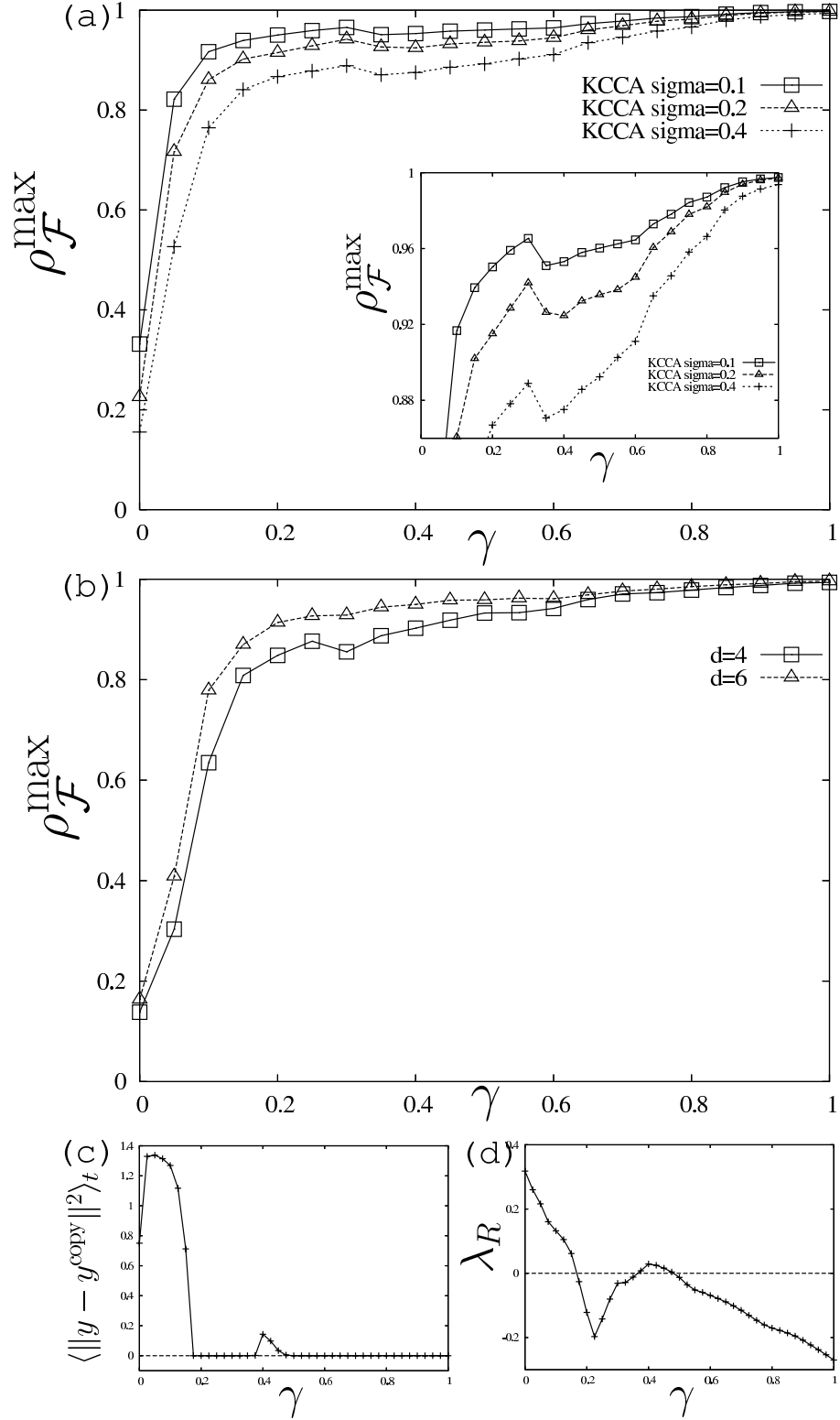}
\caption{(a) The canonical correlation coefficient $\rho^{\max}_{\cal F}$ of the Kernel CCA as a function of the coupling strength $\gamma$ for three different values of $\sigma$ with $\kappa = 0.1$  and $N=2\times 10^3$.  Inset An enlargement of (a).  (b) The index $\rho^{\max}_{\cal F}$ vs. $\gamma$  for the embedding dimension $d=4$ and $d=6$  with $\sigma=1.0, \kappa=0.1$, and $N=2\times 10^3$.  The delay time $l$ for embedding is set to $1$. (c) The time average of the synchronization error between the response system and its identical copy as a function of $\gamma$. (d) The maximal Lyapunov exponent $\lambda_R$ of the response system as a function of  $\gamma$.}
\label{revise_figure2}
\end{figure}

We apply the Kernel CCA to the cases shown in figures~\ref{revise_figure1} (a) and (b).
Here, we employ a Gaussian kernel with $\sigma = 0.1$ and $\kappa$ is set to $0.3$.
We prepare an orbit with length $N=2\times 10^3$ as training data.
As already remarked, each of variables $x_1$, $x_2$, $y_1$, and $y_2$ is normalized within the unit interval $[0,1]$.
Figures~\ref{revise_figure1} (c) and (d) illustrate results of the Kernel CCA.
Figure~\ref{revise_figure1} (d) shows that GS is clearly identified as a cloud of points along the diagonal on the plane of the canonical variates $u$ and $v$.
On the other hand, when $X$ and $Y$ are independent, the correlation between the canonical variates is very weak as shown in figure~\ref{revise_figure1} (c). 
Figure~\ref{revise_figure2} (a) shows the dependence of the canonical correlation coefficient $\rho^{\max}_{\cal F}$ on the coupling strength $\gamma$.
The rapid increase of $\rho^{\max}_{\cal F}$ against $\gamma$ in an interval $0<\gamma<0.17$ is indicated.

We will show that the Kernel CCA can be used with time-delay embedding scheme for time series.
In time-delay embedding scheme, a pair of $d$-dimensional vectors $(x_1 (t), x_1 (t-l), ..., x_1 (t-(d-1)l)$ and $(y_1 (t), y_1 (t-l), ..., y_1 (t-(d-1)l)$,  where $l$ denotes the delay time, is used as a sample of training data. 
Figure~\ref{revise_figure2} (b) shows the graph of $\rho^{\max}_{\cal F}$ vs. $\gamma$ for two  different values of the embedding dimension $d$.
Shapes of the graphs shown in figure~\ref{revise_figure2} (b) does not change significantly  compared to those shown in figure~\ref{revise_figure2} (a).  
This result indicates that the Kernel CCA works for data obtained by using the time-delay embedding scheme. 

%
\subsection{Comparison with A Conventional Method}
In \cite{Koca96,Abar96}, an identification of GS based on the occurrence of the complete synchronization between the response system and its identical copy is proposed.
For the coupled H\'{e}non maps (\ref{Henon_1}) and (\ref{Henon_2}),  we investigate the synchronization errors $\langle \Vert y-y^{\rm copy}\Vert^2\rangle_t$  between the system $Y$ and its identical copy $Y'$ with variables $y^{\rm copy} = (y_1^{\rm copy}, y_2^{\rm copy})$. 
The result is shown in figure~\ref{revise_figure2} (c).
Here, the index $\langle \Vert y-y^{\rm copy}\Vert^2\rangle_t$ between $Y$ and $Y'$ is measured by the average of $10^6$ time steps and plotted as a function of $\gamma$.
The index $\langle \Vert y-y^{\rm copy}\Vert^2\rangle_t$ decreases with increasing $\gamma$ and becomes zero at $\gamma\sim 0.17$.
%
Our results shown in figures~\ref{revise_figure2} (a) and (b) are consistent with this result. 

We also investigate the maximal Lyapunov exponent  $\lambda_R$ of the state $y(t)=y^{\rm copy} (t)$.
The result is shown in figure~\ref{revise_figure2} (d).   
The index $\lambda_R$ is determined from the first eigenvalue of 
the product of the Jacobian matrix of (\ref{Henon_2}) with respect to the variables $y$ 
,  and an orbit with length $10^6$ is used for its numerical evaluation.   
There is an interval where $\lambda_R$ changes nonmonotonically against $\gamma$.
Such nonmonotonic change is also observed in the graph of $\rho^{\max}_{\cal F}$ vs. $\gamma$ as shown in the inset of figure~\ref{revise_figure2} (a).  
These results tell us that $\rho^{\max}_{\cal F}$ defined by the Kernel CCA can characterize subtle change as well as global tendency of GS. 

%

\subsection{Influence of Noise and Sample Size}
\begin{figure}[hptb]
\includegraphics[width=12cm]{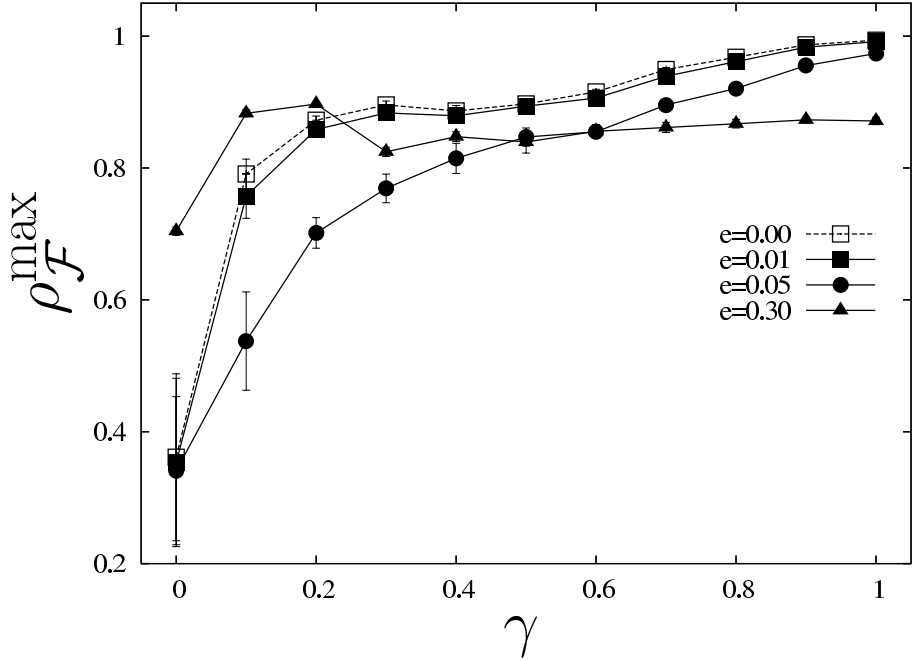}
\caption{(a) Influence of the observational noise on the index $\rho^{\max}_{\cal F}$ as a    function of $\gamma$ with $\sigma=0.4, \kappa=0.1$, and $N=10^3$. The average and the standard deviation of $\rho^{\max}_{\cal F}$ over $20$ realizations are plotted as symbols and vertical bars, respectively.}
\label{revise_figure3}
\end{figure}

We consider how the performance of the Kernel CCA is influenced by the introduction of the observational noise and the change of the size of training data.  
First we consider the influence of noise.
In numerical simulation, for each variable of (\ref{Henon_1}) and (\ref{Henon_2})  normalized within the interval $[0,1]$, Gaussian random numbers with the mean zero and the standard deviation $g$ are added as the observational noise. 
Results are shown in figure~\ref{revise_figure3}. 
For a low noise level ($g=0.01$),  the graph of $\rho^{\max}_{\cal F}$ vs. $\gamma$ is almost same as that for the noise free case.
For higher noise levels ($g= 0.05$),  there is  a moderate decrease of $\rho^{\max}_{\cal F}$  in the whole interval of $\gamma$, however, the global tendency of the graph of $\rho^{\max}_{\cal F}$ vs. $\gamma$ is not lost.
The proposed approach is fairly robust against noise except for extremely high noise level ($g=0.3$).

\begin{figure}[hptb]
\includegraphics[width=12cm]{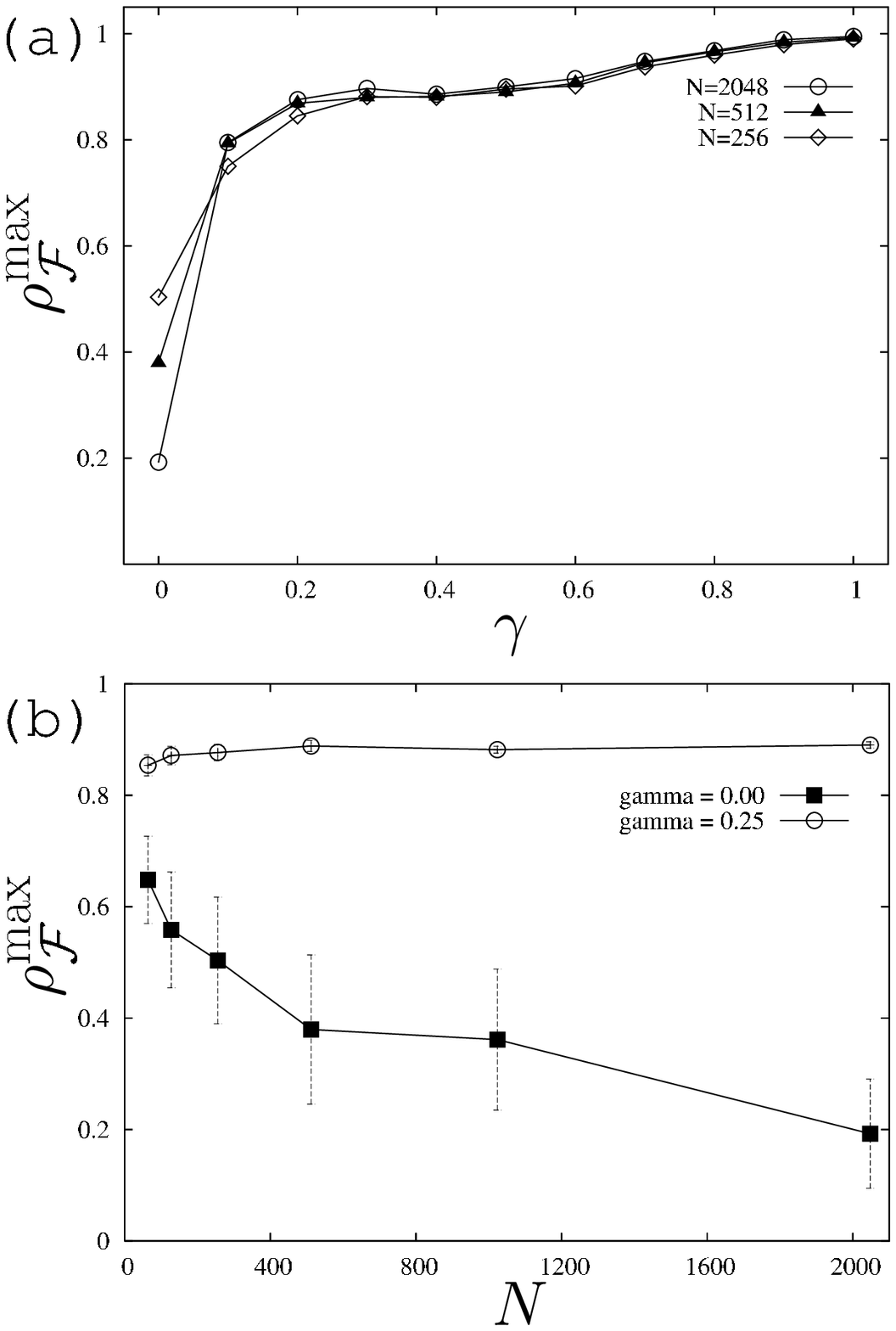}
\caption{(a) Dependence of the index $\rho^{\max}_{\cal F}$ on $\gamma$ for three different values of the size $N$ of training data with $\sigma=0.4, \kappa=0.1$.
(b) Dependence of the index $\rho^{\max}_{\cal F}$ on $N$ for $\gamma=0$ and $\gamma = 0.25$ with $\sigma=0.4, \kappa=0.1$. 
The average of $\rho^{\max}_{\cal F}$ over $20$ realizations are plotted as symbols in (a), and in addition to the average, the standard deviation is also plotted  as vertical bars in (b). }
\label{revise_figure4}
\end{figure}

Second, we study the influence of the size of training data.
Figure~\ref{revise_figure4} (a) shows the average of $\rho^{\max}_{\cal F}$ over $20$ realizations as a function of $\gamma$ for several different values of the size $N$.
With the exception of $\gamma = 0$,  the graph of $\rho^{\max}_{\cal F}$ vs. $\gamma$ does not depend much on $N$.
The result shows that the Kernel CCA works even with relatively small size of training data.
For the cases of $\gamma = 0$ and $\gamma = 0.25$, the graphs of $\rho^{\max}_{\cal F}$ vs. $N$ are shown in figure~\ref{revise_figure4} (b). 
Here, the vertical bars denote the corresponding standard deviation.  
The average of  $\rho^{\max}_{\cal F}$ for $\gamma = 0$ increases monotonically with decreasing $N$ whereas the one for $\gamma = 0.25$ does not almost change against $N$.
The result with $\gamma=0$ suggests that a proper choice of the kernel parameter for a given data is important for obtaining correct results. 
This issue is discussed in Section 6. 

\subsection{Assessing Sensitivity of Kernel CCA to Nonlinear Structure: Surrogate Data Analysis}
In order to investigate the ability of the Kernel CCA to nonlinear dependence between two systems (\ref{Henon_1}) and (\ref{Henon_2}) , we use a method of surrogation\cite{Schr00}. 
Multivariate surrogate data are generated as follows: 
first, the Fourier transform of the time series is calculated for each of variables, then the {\it common} random numbers are added to the phase variables, and finally  the inverse Fourier transform is applied.
The resulting multivariate time series have the same power spectra and cross spectra as those of the original time series.
By changing random numbers added to the phases, an arbitrary number of  different time series which preserve the linear properties of the original is obtained.  
See papers \cite{Pric94,Schr96,Schr00} for technical details. 

In numerical simulation, 19 realizations of the surrogate data for the time series $\{(x_1 (t), y_1 (t))\}$ of (\ref{Henon_1}) and (\ref{Henon_2}) are prepared by using  the TISEAN package\cite{Hegg99,Schr00}. 
We take $d=4$ and $l=1$ for time-delay embedding. 

\begin{figure}[hptb]
\includegraphics[width=12cm]{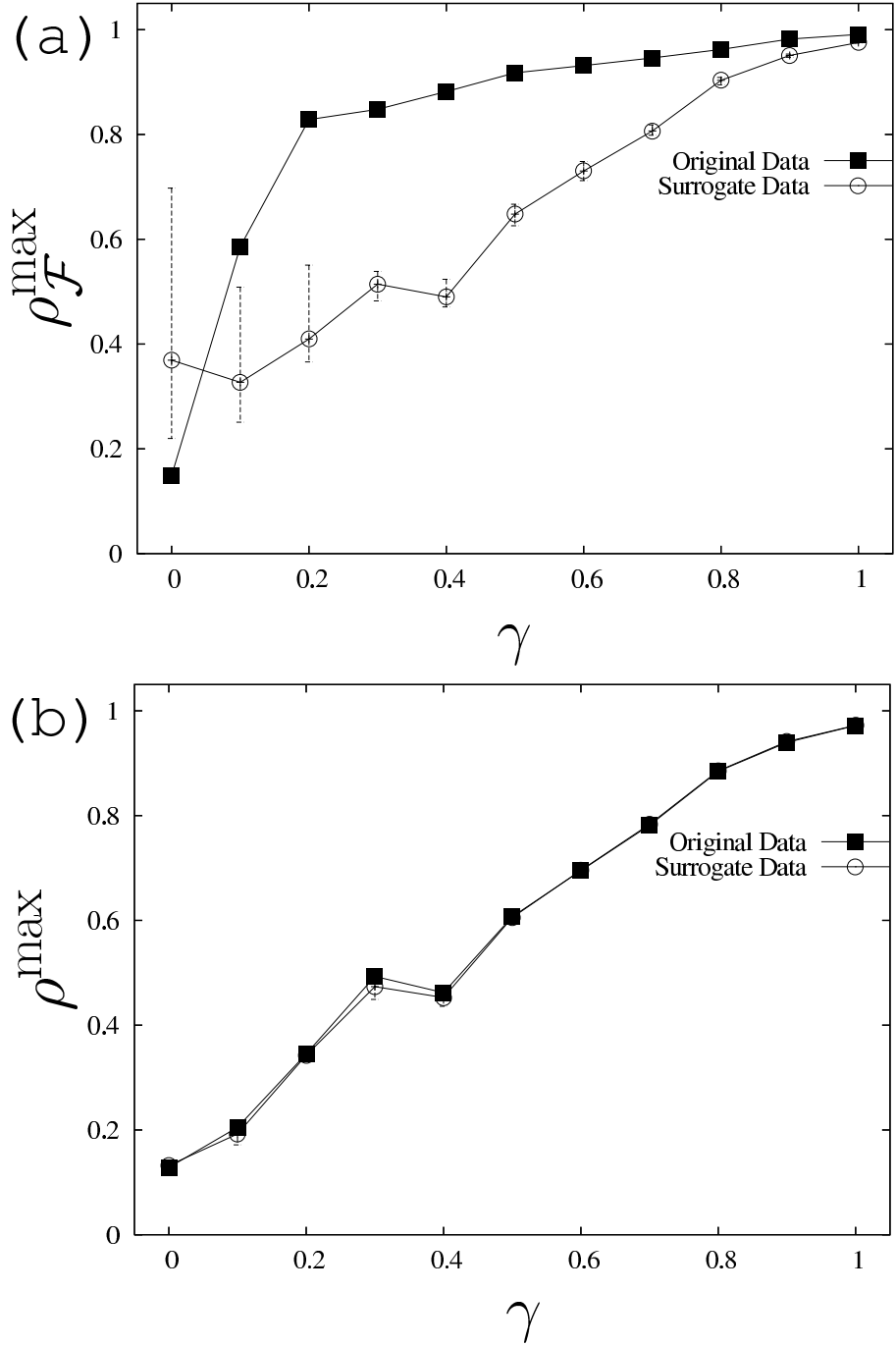}
\caption{The canonical correlation coefficients $\rho^{\max}_{\cal F}$ of the Kernel CCA (a) and $\rho^{\max}$ of the linear CCA (b) as functions of $\gamma$ for the original data and the surrogate data. For the surrogate data, the average, and the maximal and minival values of  $\rho^{\max}_{\cal F}$  over  $19$ realizations are plotted as symbols, and both edges of vertical bars, respectively. Parameters are $d=4 ,l =1, \sigma=1.0, \kappa=0.1$, and $N=10^3$ in (a) and $d=4, l=1$ and $N=10^3$ in (b).}
\label{revise_figure5}
\end{figure}

In figure~\ref{revise_figure5} (a), the index $\rho^{\max}_{\cal F}$ defined by the Kernel CCA for the original data and that for the surrogate data as functions of $\gamma$ are shown. 
For the surrogate data, the average over 19 realizations is plotted as a function of $\gamma$, and  the corresponding maximal and minimal values  are also shown as  the both edges of vertical bars. 
For both of the original and the surrogate data,  the index $\rho^{\max}_{\cal F}$ increases with increasing $\gamma$.
Except for $\gamma = 0$ and $\gamma \sim 1$, however, the value of $\rho^{\max}_{\cal F}$  for the original data is significantly higher than that for the surrogate data.
For larger values of $\gamma$, $\rho^{\max}_{\cal F}$ for the surrogate data increases monotonically, and $\rho^{\max}_{\cal F} \to 1$ as $\gamma \to 1$.
This coincides with the fact that the attractor of the systems (\ref{Henon_1}) and (\ref{Henon_2}) is located around the plane $x_1=y_1$ and $x_2=y_2$  and the relation between two systems becomes almost  linear one.
The results suggest that the Kernel CCA is sensitive to nonlinearity of the dependence between two systems.

We also investigate the performance of the linear CCA in the same way and results are shown in figure~\ref{revise_figure5} (b). 
%
In this case, the difference between the maximal canonical correlation coefficients  $\rho^{\max}$ of the linear CCA for the original data and the one  for the surrogate data is not significant for any value of $\gamma$. 
This indicates that the linear CCA can detect only the linear dependence between two systems. 

\subsection{The Regularization Parameter $\kappa$}
\begin{figure}[hptb]
\includegraphics[width=10cm]{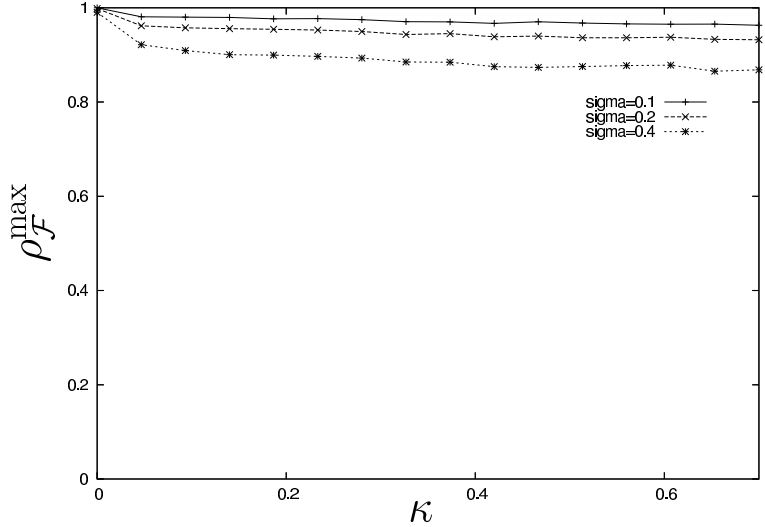}
\caption{The index $\rho_{\cal F}^{\max}$ as a function of the regularization parameter $\kappa$ for three different values of $\sigma$. $\gamma=0.3$ and $N=3\times 10^2$.}
\label{revise_figure6}
\end{figure}

As mentioned in the preceding section, the regularization terms are required for nontrivial results. 
Figure~\ref{revise_figure6} shows the dependence of $\rho^{\max}_{\cal F}$ on the regularization parameter $\kappa$ for three different values of $\sigma$. 
Although $\rho^{\max}_{\cal F}\to 1$ for too small $\kappa$, the value of $\rho^{\max}_{\cal F}$ decreases gradually and does not depend on the precise value of $\kappa$. 

\section{Other  Examples}
In order to illustrate the capability of the Kernel CCA in more complicated situations, we add  the following three examples. 
\subsection{Coupled R\"{o}ssler-Lorenz Systems}
First, we consider GS in a Lorenz system driven by a R\"{o}ssler system\cite{Abar96}: 
\begin{eqnarray}
\label{Ross2}
&& X:\left\{
\begin{array}{lll}
\dot{x_1} &= - (x_2+x_3) \\
\dot{x_2} &=  x_1 + 0.2 x_2 \\
\dot{x_3} &=  0.2 + x_3(x_1-5.7), 
\end{array}
\right.
\\
\label{Lore}
&& Y:\left\{
\begin{array}{lll}
\dot{y_1} &=  16(y_2 - y_1) - \gamma (y_1 - x_1)\\
\dot{y_2} &=  -y_1 y_3 + 45.92 y_1 - y_2 \\
\dot{y_3} &=  y_1y_2 - 4y_3.
\end{array}
\right.
\end{eqnarray}
The coupling term is introduced in the first equation of (\ref{Lore}), and $\gamma$ is its strength. 

\begin{figure}[hptb]
\includegraphics[width=12cm]{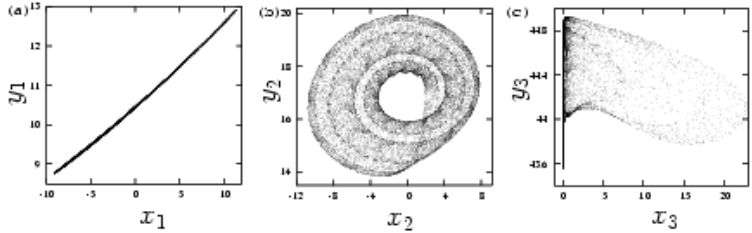}
\caption{Projections of the strange attractor of the coupled R\"{o}ssler and Lorenz systems with $\gamma = 10$ onto the planes of $(x_1, y_1)$ in (a), $(x_2, y_2)$  in (b), and $(x_3, y_3)$  in (c). }
\label{revise_figure7}
\end{figure}

We confirm that there is a sharp transition of GS at $\gamma\sim 4.8$ by investigating the long time average of the synchronization error between the system (\ref{Lore}) and its identical copy driven by the common signal $x_1 (t)$ of the system (\ref{Ross2}).  
There coexist two attractors in the state space after the transition of GS and we choose  one of them. 
Figures~\ref{revise_figure7} (a) -- (c) show the projections of the strange attractor of the system (\ref{Ross2}) and (\ref{Lore}) with $\gamma = 10$ onto the planes of $(x_1, y_1)$, $(x_2,y_2)$, and $(x_3, y_3)$ respectively.

\begin{figure}[hptb]
\includegraphics[width=12cm]{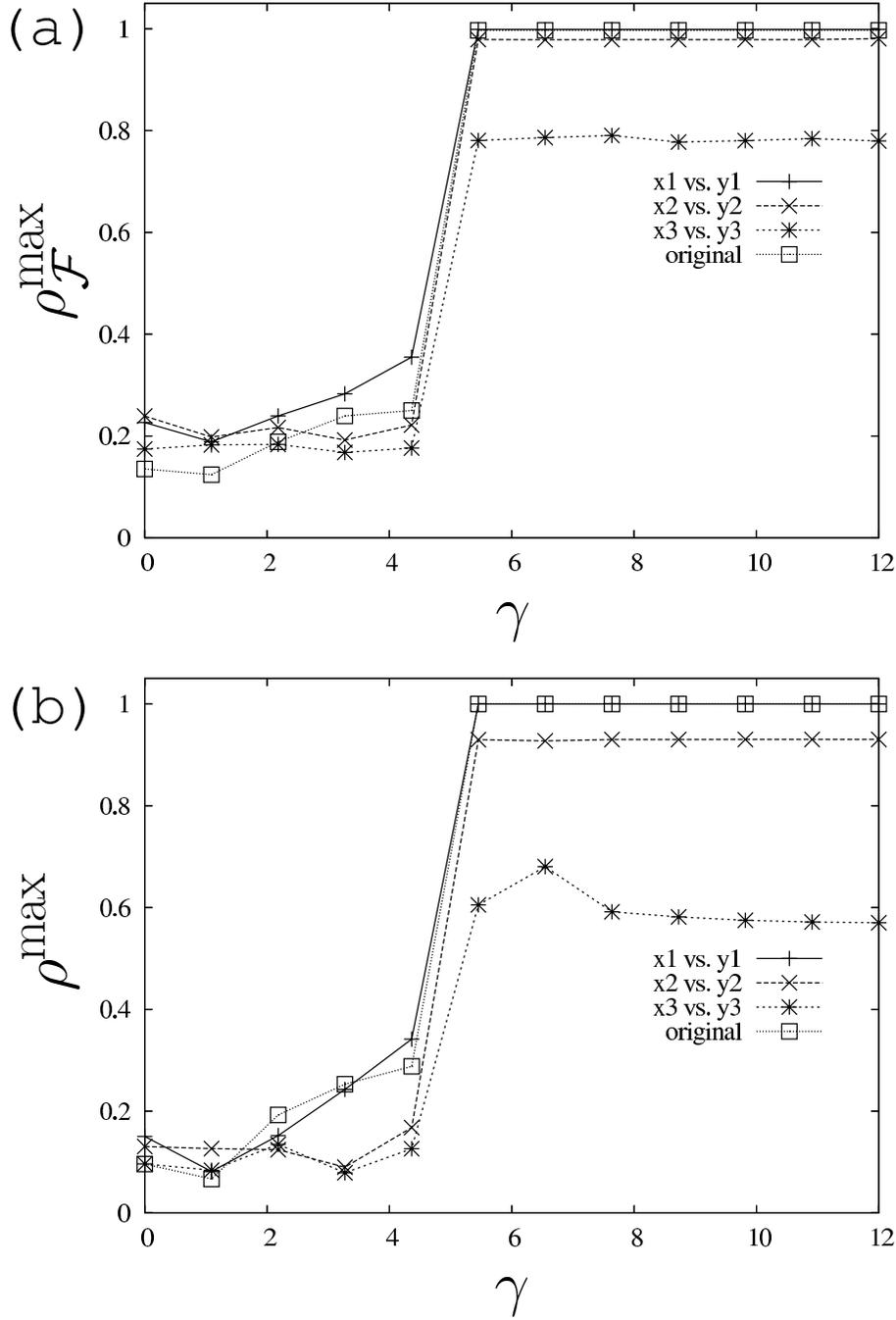}
\caption{(a) The canonical correlation coefficient $\rho^{\max}_{\cal F}$ of the Kernel CCA as a function of the coupling strength $\gamma$ for different pairs of variables with $N=10^3, d=6, l=2.5, \sigma=1.5$ and $\kappa = 0.1$.
(b) The canonical correlation coefficient $\rho^{\max}$ of the linear  CCA as a function of the coupling strength $\gamma$ for different pairs of variables with $N=10^3, d=6$, and $l=2.5$.
The results for the case where the original state variables $(x_1, x_2, x_3)$ and $(y_1, y_2, y_3)$ are used as training data are also shown.}
\label{revise_figure8}
\end{figure}

For each of pairs shown in figures~\ref{revise_figure7} (a) -- (c), the time series is embedded as points in a $d$-dimensional state space, and we apply the Kernel CCA. 
We set the embedding dimension $d=6$ and the delay time $l=2.5$.
The size of training data is $N=10^3$. 
Results are shown in figure~\ref{revise_figure8} (a). 
All indices $\rho^{\max}_{\cal F}$ shown in  figure~\ref{revise_figure8} (a) take large values for $\gamma \gtrsim 4.8$.
The value $\gamma \sim 4.8$ agrees with the transition point of GS. 
In figure~\ref{revise_figure8} (b), results of the linear CCA applied to the same data are also shown.
The indices  $\rho^{\max}$ of the linear CCA for the cases of $(x_1$ vs. $y_1)$ and $(x_2$ vs. $y_2)$  take large values after the GS transition as well as the indices $\rho^{\max}_{\cal F}$  obtained by  the Kernel CCA.
For the case of $(x_3$ vs. $y_3)$, however,  $\rho^{\max}$  of the linear CCA is smaller than  $\rho^{\max}_{\cal F}$ of the Kernel CCA.
This result suggests that the Kernel CCA outperforms the linear CCA when the relation between two observed time series has high nonlinearity. 
 
\subsection{Neural Spike Trains Modulated by Chaotic Inputs}
Second, we analyze the following FitzHugh-Nagumo (FHN) neuron model modulated by chaotic dynamics of the R\"{o}ssler model:   
\begin{eqnarray}
\label{Ross}
&& X:\left\{
\begin{array}{lll}
\dot{x_1} &= -\tau (x_2+x_3) \\
\dot{x_2} &= \tau (x_1 + 0.36 x_2) \\
\dot{x_3} &= \tau (0.4 + x_3(x_1-4.5)), 
\end{array}
\right.
\\
\label{FHN}
&& Y:\left\{
\begin{array}{lll}
\dot{y_1}   &=   \{-y_1(y_1-0.5)(y_1-1) - y_2 +  S(t)\}/0.02\\
\dot{y_2}   &=  y_1 - y_2 - 0.15,
\end{array}
\right.
\end{eqnarray}
where the term
\begin{eqnarray}
\label{input}
S(t) = 0.23 + 0.0075  x_1(t)
\end{eqnarray}
defines a chaotic inputs to the neuron,  and $\tau$ is a parameter that controls the dominant time scale of the  R\"{o}ssler dynamics.   
The system composed of  (\ref{Ross}) and (\ref{FHN}) has been investigated from the viewpoint of the problem whether the information on the input signal can be decoded from the output interspike intervals (ISIs) generated by a neuron or not\cite{Saue94,Raci97,Han02}.    
Here, we focus on the relation between the input chaotic stimulus and the output ISIs from the viewpoint of GS between two oscillators with different dynamics.

We define the $i$-th ISI as $s_i = t_i - t_{i-1}$ where $t_i$ is onset time of the $i$-th spike defined as the time when the variable $y_1$ makes upward crossing over some fixed threshold $y_{\theta}$.
The value of  $y_{\theta}$ is set to $0.7$ here.
We also define the chaotic stimulus associated with the $i$-th ISI as $r_i  \equiv x_1(t_i)$, which is the value of $x_1$ in   (\ref{Ross}) at $t_i$.
By using the delay embedding scheme, we transform the time series $\{r_i\}$ and $\{s_i\}$ into the state points $\{(r_i, r_{i-1}, ..., r_{i-d_X+1})\}$ and $\{(s_i, s_{i-1}, ..., s_{i-d_Y+1})\}$ in $d_X$ and $d_Y$-dimensional state spaces, respectively.
We set $d_X=d_Y=3$ here and apply the Kernel CCA to these data sets. 

\begin{figure}[hptb]
\includegraphics[width=12cm]{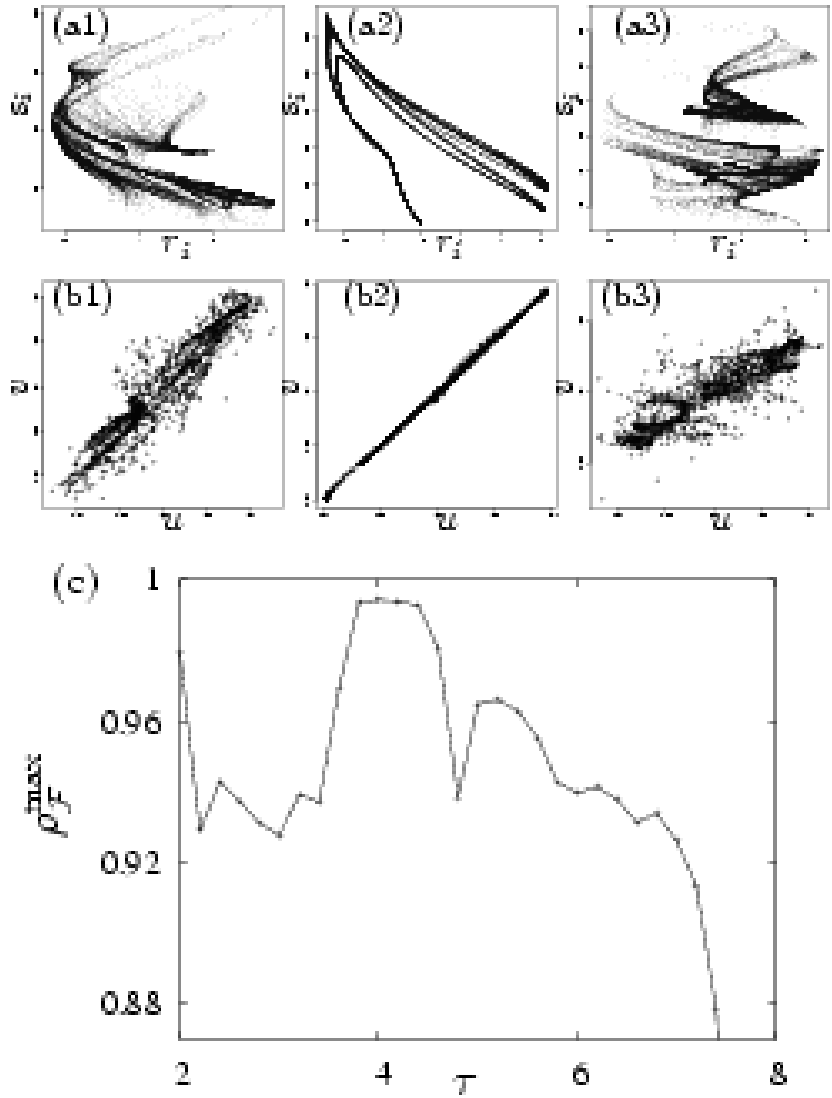}
\caption{(a1 -- a3) Scatter plots of $r_i$ and $s_i$ of  (\ref{Ross}) and (\ref{FHN}) with $\tau = 2.7$ in (a1), $\tau=4.0$ in (a2), and $\tau=7.5$ in (a3).  (b1 -- b3) Scatter plots of canonical variates $u$ and $v$ of the Kernel CCA with $\tau=2.7$ in (b1), $\tau=4.0$ in (b2), and $\tau=7.5$ in (b3). $\sigma=0.5$, $\kappa=0.1$, and $N=2\times 10^3$. (c) The canonical correlation coefficient $\rho^{\max}_{\cal F}$ of the Kernel CCA as a function of $\tau$ with $\sigma=0.5$, $\kappa=0.1$, and $N=2\times 10^3$. }
\label{revise_figure9}
\end{figure}

In figures~\ref{revise_figure9} (a1)--(a3), significant nonlinear dependences between the input chaos and the output ISI are observed in scatter plots of $(r_i, s_i)$.
Figures~\ref{revise_figure9} (b1)--(b3) also show scatter plots of the canonical variates of the Kernel CCA associated with figures~\ref{revise_figure9} (a1)--(a3). 
It is easy to see that the correlation between canonical variates $u$ and $v$ shown in figures~\ref{revise_figure9} (b1)--(b3) correspond to the complexity of nonlinear dependence between $r_i$ and $x_i$ shown in figures~\ref{revise_figure9} (a1)--(a3).

The relation between $r_i$ and $s_i$ changes according to the value of the control parameter $\tau$.  
In figure~\ref{revise_figure9} (c), the canonical correlation coefficient $\rho^{\max}_{\cal F}$ is plotted as a function of $\tau$, which visualizes the change of the input-output relation between two systems of  (\ref{Ross}) and (\ref{FHN}). 
The index $\rho^{\max}_{\cal F}$ changes nonmonotonically with the increase of $\tau$, and there is a regime around $\tau\sim 4$ where the value of  $\rho^{\max}_{\cal F}$ is large. 
In addition to GS, chaotic phase synchronization (CPS)~\cite{Rose96} occurs between two systems in this regime~\cite{Han02}. 
The increase of $\rho^{\max}_{\cal F}$ in this regime can be attributed to the occurrence of CPS. 

\subsection{Bidirectionally Coupled Systems}
The notion of GS is not resricted to unidirectinally coupled systems.
In~\cite{Zhen02,Osip03}, the occurrence of GS for bidirectionally coupled systems is also discussed.
As an example of GS in bidirectinally coupled systems, we consider the following coupled Lorenz-R\"{o}ssler systems~\cite{Zhen02}: 
 \begin{eqnarray}
\label{Bi_Lore}
&& X:\left\{
\begin{array}{lll}
\dot{x_1} &= 10 (x_2 - x_1) + \gamma (y_1 - x_1) \\
\dot{x_2} &=  35 x_1 - x_2 -x_1 x_3 \\
\dot{x_3} &=  -(8/3) x_3 + x_1 x_2, 
\end{array}
\right.
\\
\label{Bi_Ross}
&& Y:\left\{
\begin{array}{lll}
\dot{y_1} &=  5.5 y_2 - y_3  + \gamma (x_1 - y_1)\\
\dot{y_2} &=  5.5 y_1 + 0.165 y_2\\
\dot{y_3} &=  0.2 + y_3 (y_1 - 10).
\end{array}
\right.
\end{eqnarray}
A mutual interaction between a Lorenz system (\ref{Bi_Lore}) and a R\"{o}ssler system (\ref{Bi_Ross}) is introduced as  diffusion terms in the first equations of both (\ref{Bi_Lore}) and (\ref{Bi_Ross}), and $\gamma$ is its strength.
In~\cite{Zhen02}, Zheng et al. try to define GS in bidirectionally coupled systems by considering identical copies $X'$ and $Y'$ of the systems $X$ and $Y$.
Using the approach in \cite{Zhen02}, two transitions of GS are found in the systems (\ref{Bi_Lore}) and (\ref{Bi_Ross}). 
At $\gamma\sim 1.2$, the R\"{o}ssler system $Y$ is entrained by the Lorenz system $X$ in the sense that orbits of the systems $Y$ and $Y'$ completely coincide with each other by receiving the common signal of the system $X$.
With the further increase of $\gamma$, the system $X$ is also entrained by the system $Y$ at $\gamma\sim 12.3$ which means that the complete synchronization between $X$ and $X'$ also occurs.

\begin{figure}[hptb]
\includegraphics[width=12cm]{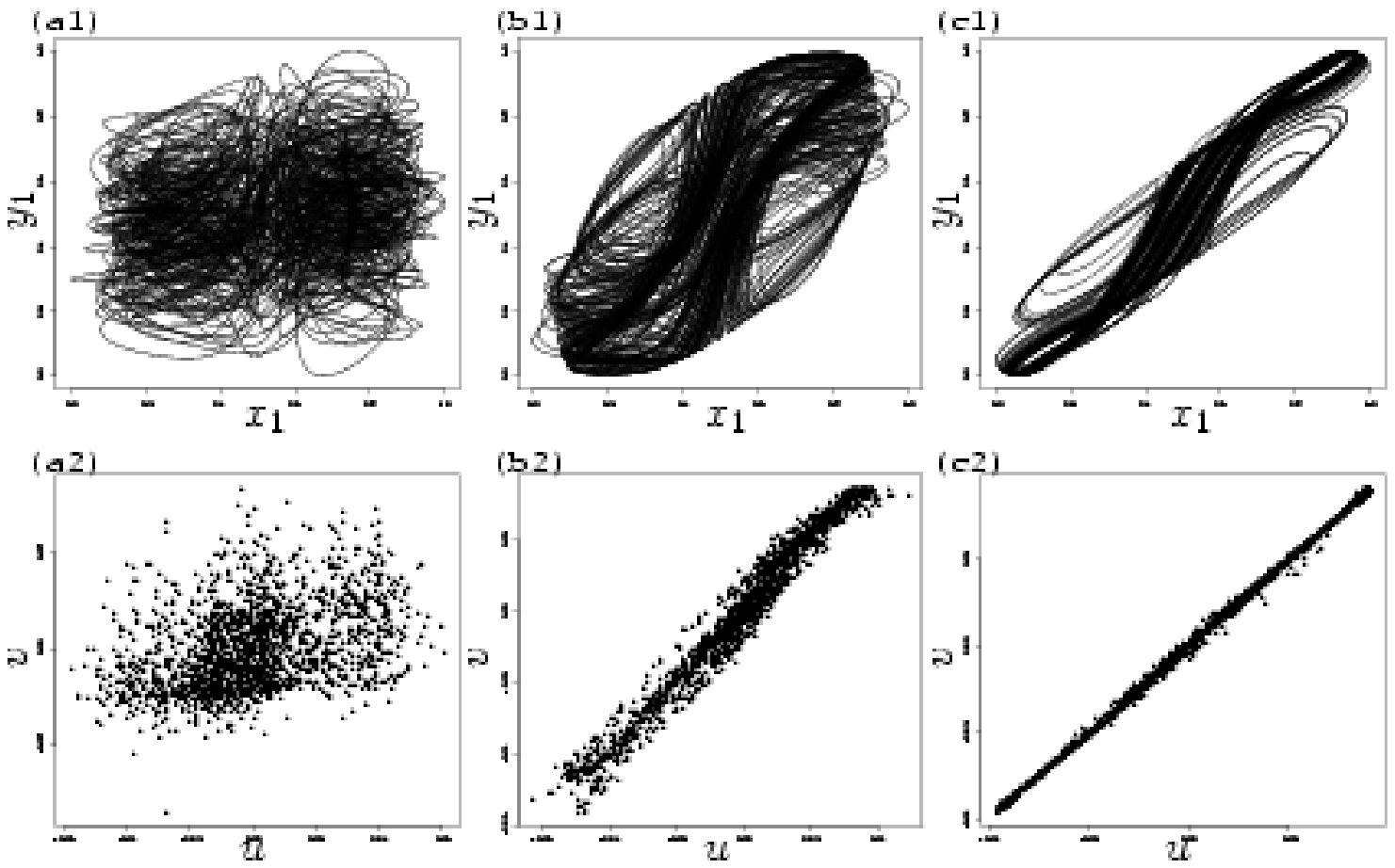}
\caption{(a1, b1, c1) Projections of attractors of the bidirectionally coupled Lorenz and R\"{o}ssler systems onto the plane of $(x_1, y_1)$ with $\gamma=0.4$ in (a1),  $\gamma=4.0$ in (b1) $\gamma=13.0$ in (c1). (a2, b2, c2) Scatter plots of the canonical covariates of the Kernel CCA  with $\gamma=0.4$ in (a1),  $\gamma=4.0$ in (b1) $\gamma=13.0$ in (c1). Parameters are $\sigma=1.0, \kappa =0.1$, and $N=2\times 10^3$.}
\label{amend_revise1}
\end{figure}

\begin{figure}[hptb]
\includegraphics[width=12cm]{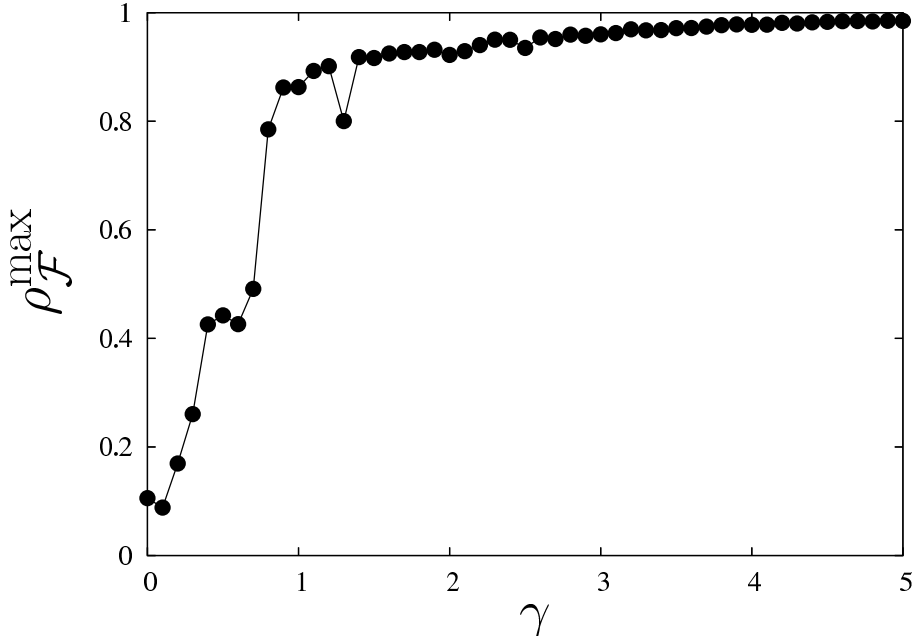}
\caption{The index $\rho_{\cal F}^{\max}$ of the Kernel CCA as a function of the coupling strength $\gamma$ with $\sigma=1.0, \kappa =0.1$, and $N=2\times 10^3$.}
\label{amend_revise2}
\end{figure}

We apply the Kernel CCA to the systems (\ref{Bi_Lore}) and (\ref{Bi_Ross}) and results are shown in figures~\ref{amend_revise1} and \ref{amend_revise2}.
Figure~\ref{amend_revise1} (a1), (b1), and (c1) show the projections of attractors of the systems (\ref{Bi_Lore}) and (\ref{Bi_Ross}) with $\gamma = 0.4$, $4.0$, and $13.0$ onto the $(x_1, y_1)$ planes, respectively. 
Figures~\ref{amend_revise1}  (a2), (b2) and (c2) also show the corresponding results of the Kernel CCA.  
Here, we use a Gaussian kernel with $\sigma=1.0$.
The state variables $\{(x_1 (t_i), x_2(t_i), x_3(t_i)), (y_1(t_i), y_2(t_i), y_3(t_i))\}$ where $t_i = i\Delta t, i=1,...,N, \Delta t = 5.0, N=2\times 10^3$ are used as the training data set. 
It is observed that the state of GS is clearly identified as the high linear correlation between canonical variates $u$ and $v$ of the Kernel CCA.
Application of the Kernel CCA to the bidirectionally coupled systems is straightforward while the approach of  \cite{Zhen02} requires rather subtle procedures.
Figure~\ref{amend_revise2}  shows the index $\rho_{\cal F}^{\max}$ as a function of the coupling strength $\gamma$. 
There is a rapid increase of  $\rho_{\cal F}^{\max}$ against $\gamma$ in an interval $0<\gamma<1.2$.
This result is consistent with the first transition of GS defined in \cite{Zhen02}.
The second transntion of GS at $\gamma\sim 12.3$ defined in \cite{Zhen02} is not seen in the graph of $\rho_{\cal F}^{\max}$ vs. $\gamma$.
A reason is that the value of $\rho_{\cal F}^{\max}$ already becomes nearly one at $\gamma\sim 5$.
Another reason is that by definition, the proposed approach based on the Kernel CCA is insensitive to the directionality of synchronization.
It will be interesting to study  modifications of the approach which can deal with  the directionality of synchronization.

\section{Nonstationary Change of Coupling Strength}
So far, we have focused on the dependence of the first eigenvalue $\rho^{\max}_{\cal F}$ on the coupling strength $\gamma$. 
We turn our attention to the eigenvector ${}^t (\alpha, \beta)$ in  (\ref{KCCA_2})  and investigate changes of the structure of the dynamical system with the Kernel CCA. 

\begin{figure}[hptb]
\includegraphics[width=12cm]{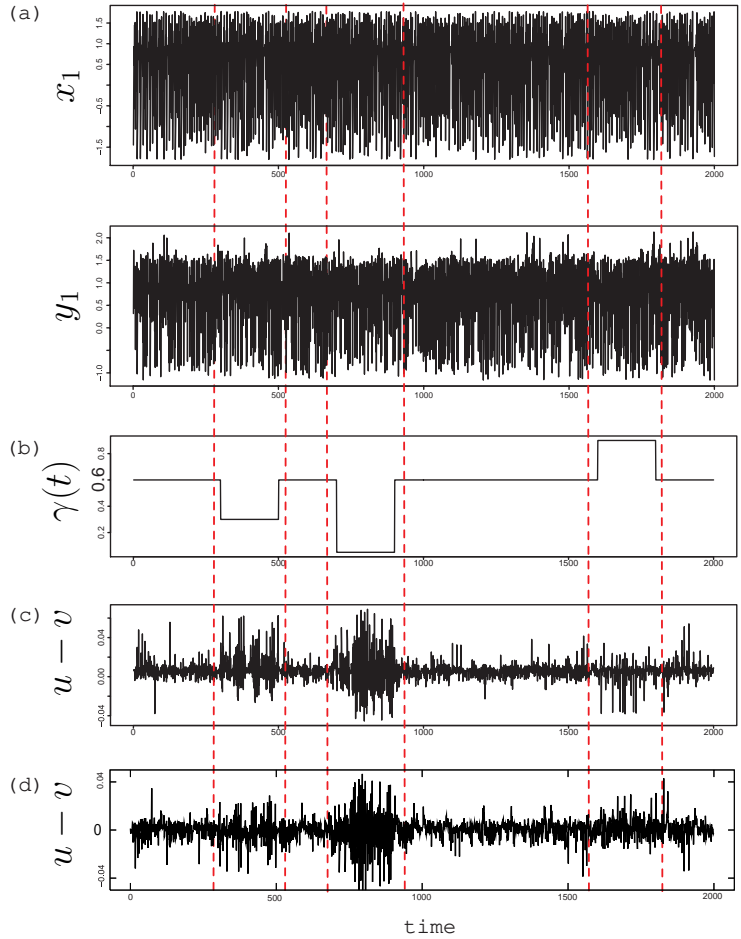}
\caption{(a) Time series of the variables $x_1$ and $y_1$ of the coupled H\'{e}non maps when the value of $\gamma$ changes temporally as shown in (b). (c, d) Time series of the difference of the canonical variates of the Kernel CCA with $\sigma=0.1$ and $\kappa=0.1$.}
\label{revise_figure10}
\end{figure}

As an illustration, let us consider the problem of extracting nonstationary changes of the coupling strength $\gamma$ from time series data generated by the coupled H\'{e}non maps.
An example of such changes is shown in figure~\ref{revise_figure10} (b).  
When the value of $\gamma$ varies from $\gamma_0$, an orbit leaves from the synchronization manifold ${\cal M}$ with $\gamma_0$, and is attracted again to it when the value of $\gamma$ is restored to $\gamma_0$. 
As shown in figure~\ref{revise_figure10} (a), the time series of  the original variables $(x_1, y_1)$ does not  tell us whether the orbit is lying on ${\cal M}$ or not at a given time $t$. 
By using the Kernel CCA, a deviation of the orbit from ${\cal M}$ will be detected as a large value of $|f(x) - g(y)|$, where  $f(\cdot)$ and $g(\cdot)$ are nonlinear transformations determined by the first eigenvector ${}^t(\alpha, \beta)$.

Numerical experiments are performed with $\gamma_0 = 0.6$ in two different conditions, and results are shown in figures~\ref{revise_figure10} (c) and (d), respectively.
Figure~\ref{revise_figure10} (c) shows the case where  $f(\cdot)$ and $g(\cdot)$ are estimated from an orbit with $\gamma=\gamma_0$, which is  prepared separately from the orbit to be analyzed.  
In figure~\ref{revise_figure10} (c), time intervals where the values of $\gamma$ are different from $\gamma_0$ is clearly detected as successive bursts in the time series of  $|f(x) - g(y)|$.

In figure~\ref{revise_figure10} (d), we show the case where the orbit to be analyzed is also used for estimating $f(\cdot)$ and $g(\cdot)$.
We see that even when estimation of nonlinear transformations 
is affected by ``noise" from the points not lying on ${\cal M}$,  successive bursts are still observed in the time series of $|f(x) - g(y)|$ except for the one in a time interval $1600 <  t<1800$.

\section{Choice of Kernel Parameters}
\begin{figure}[hptb]
\includegraphics[width=12cm]{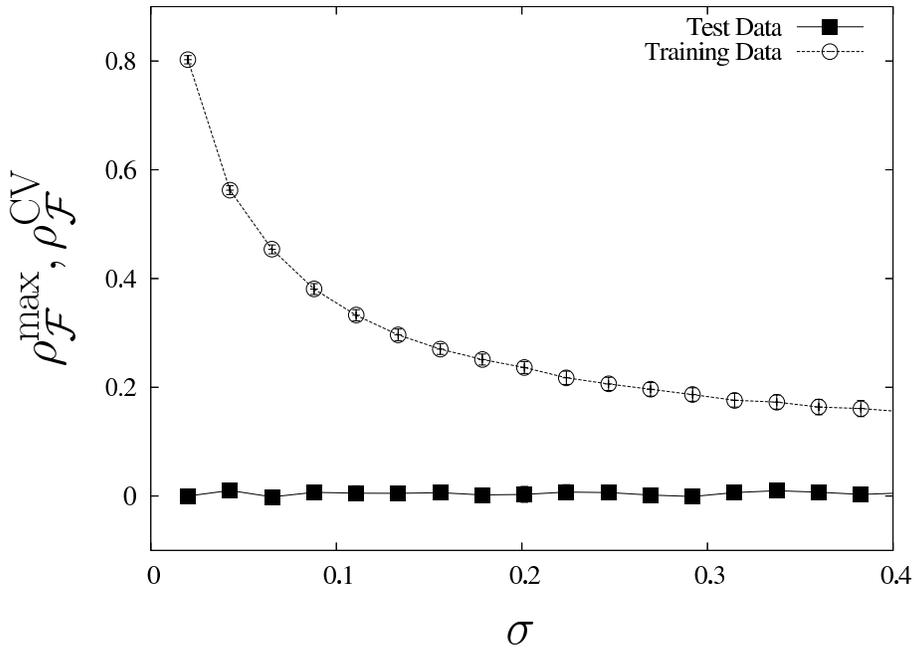}
\caption{Result of cross-validation for two uncoupled H\'{e}non maps ($\gamma = 0.0$). 
The dotted line with open circles shows $\rho_{\cal F}^{\max}$ vs. $\sigma$, and the solid line with filled squares shows $\rho_{\cal F}^{\rm CV}$ vs $\sigma$.  
We prepare 20  data sets with size $N=10^3$ for training the Kernel CCA, and a test data with size $\tilde{N}=2\times 10^3$ for cross-validation.
The average over $20$ realizations are plotted as symbols.
 The standard deviation is also plotted as vertical bars, but the ones with small values of errors are hidden by symbols. 
 $\kappa$ is set to $0.01$.
}
\label{revise_figure11}
\end{figure}

In the preceding section,  we set the value of the kernel parameter such as the width $\sigma$ of a Gaussian kernel in an ad-hoc manner.
If $\sigma$ is too small, the nonlinear transformations $f(\cdot)$ and $g(\cdot)$ in  (\ref{fg})  
cannot interpolate between data points of the training sample.
Contrary, if $\sigma$ is too large,  (\ref{fg}) cannot represent a highly nonlinear structure such as  the synchronization manifold ${\cal M}$ of GS. 
Thus a proper choice of the kernel parameter is crucial to obtain the good performance of the method. 
In this section, we discuss how the kernel parameter can be suitably chosen from a given data set.
A naive way of choosing $\sigma$ is to find the value of $\sigma$ that maximizes $\rho_{\cal F}^{\max}$. 
The dotted line with open circles in figure~\ref{revise_figure11}  shows the graph of an index $\rho_{\cal F}^{\max}$ as a function of $\sigma$ 
when two H\'{e}non maps (\ref{Henon_1}) and (\ref{Henon_2}) are uncoupled ($\gamma = 0$). 
Here, an orbit of (\ref{Henon_1}) and (\ref{Henon_2}) with length $N=10^3$ is used as a set of training data, and the average and the standard deviation over $20$ realizations are plotted as symbols and vertical bars, respectively. 
The index $\rho_{\cal F}^{\max}$ increases monotonically with the decrease of $\sigma$, and $\rho_{\cal F}^{\max} \sim 1$ is attained in the limit of $\sigma\rightarrow 0$, while there is no interaction between two systems $X$ and $Y$.
This monotonic tendency of  $\rho_{\cal F}^{\max}$ does not determine an optimal value of the kernel parameter $\sigma$. 

As a way to overcome this difficulty, the following procedure is proposed.
First, we set aside the data $\{(\tilde{x}_n, \tilde{y}_n)\}_{n=1}^{\tilde{N}}$ for assessing the performance of the Kernel CCA (we call this ``test" data) separately from the data used for training the Kernel CCA. 
Then, we estimate the nonlinear transformations $f(\cdot)$ and $g(\cdot)$ from the training data, and 
calculate the correlation coefficient $\rho_{\cal F}^{\rm CV}$  between the variables $\tilde{u}_n = f(\tilde{x}_n)$ and $\tilde{v}_n = g(\tilde{y}_n), n=1,2,...,\tilde{N}$ defined as  
\begin{eqnarray}
\label{uv_test}
\rho_{\cal F}^{\rm CV} = \frac{\langle (\tilde{u}_n - \langle \tilde{u}_n\rangle) \cdot  (\tilde{v}_n - \langle \tilde{v}_n\rangle) \rangle}{\sqrt{\langle (\tilde{u}_n - \langle \tilde{u}_n\rangle)^2} \cdot \sqrt{\langle (\tilde{v}_n - \langle \tilde{v}_n\rangle)^2}}, 
\end{eqnarray}
where $\langle \cdot \! \cdot \! \cdot\rangle$ is the empirical average over $\tilde{N}$ samples. 
This  strategy for assessing the performance of the estimated model with new data is regarded as a version of {\it cross-validation} (CV)~\cite{Ston74,Silv85,Wahb90}.

\begin{figure}[hptb]
\includegraphics[width=12cm]{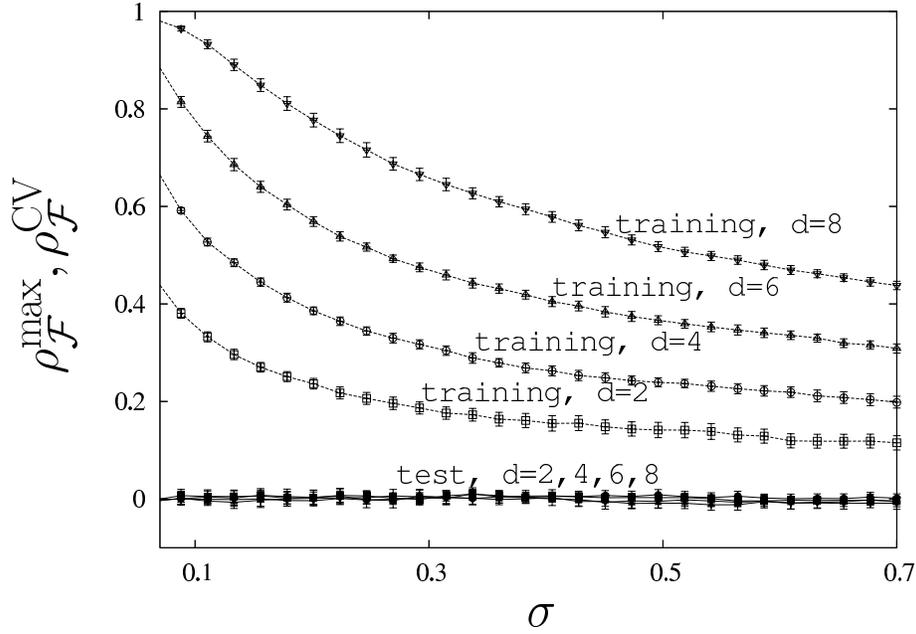}
\caption{Results of cross-validation with $\gamma = 0$ for four different values of the embedding dimension $d$. 
We take the delay time $l=1$. 
The condition for numerical experiments is the same as figure~\ref{revise_figure11}.}
\label{revise_figure12}
\end{figure}

First we check that spurious detection of synchronization can be avoided by the procedure based on CV. 
The solid line with filled squares in figure~\ref{revise_figure11}  shows the dependence of an index $\rho_{\cal F}^{\rm CV}$ on $\sigma$ with $\gamma = 0$.
When there is no interaction between two systems, the value of $\rho_{\cal F}^{\rm CV}$ is nearly equal to zero for any $\sigma$. 
%
%
Corresponding results using time-delay embedding are shown in figure~\ref{revise_figure12}. 
In figure~\ref{revise_figure12},  the dependencies of  $\rho_{\cal F}^{\max}$ and $\rho_{\cal F}^{\rm CV}$ on $\sigma$ with $\gamma=0$  are plotted for several different values of the embedding dimension $d$. 
Again, the value of $\rho_{\cal F}^{\rm CV}$ is almost equal to zero for any $\sigma$ and $d$, 
whereas the value of $\rho_{\cal F}^{\max}$ as a function of $\sigma$ increases monotonically with the increase of $d$.  
This result suggests that the procedure based on CV effectively erases spurious detection even when data is embedded in a high-dimensional  state space.

\begin{figure}[hptb]
\includegraphics[width=12cm]{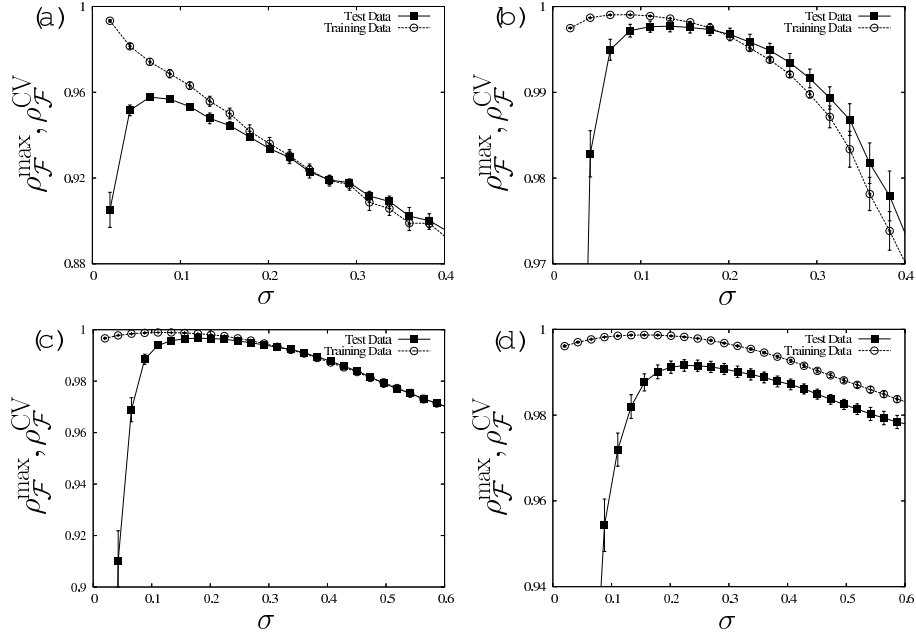}
\caption{Results of cross-validation for two coupled H\'{e}non maps with $\gamma = 0.25$. The embedding dimension $d=2$ in (a), $d=4$ in (b), $d=6$ in (c), and $d=8$ in (d), and  the delay time $l=1$. 
The condition for numerical experiments is the same as figure~\ref{revise_figure11}.}
\label{revise_figure13}
\end{figure}

Next we show that the cross validation procedure is useful for choosing optimal values of $\sigma$.
Figures~\ref{revise_figure13} shows the  values of $\rho_{\cal F}^{\max}$ and $\rho_{\cal F}^{\rm CV}$ as functions of $\sigma$ for the coupled H\'{e}non maps (\ref{Henon_1}) and (\ref{Henon_2}) with $\gamma=0.25$. 
The condition for the numerical experiment is the same as $\gamma=0$.
For all of graphs,  
the value of $\rho_{\cal F}^{\rm CV}$ with the dotted line takes its maximum at a nonzero value of $\sigma$, whereas $\rho_{\cal F}^{\max}$ increases monotonically with the decrease of $\sigma$.
This result indicates that we can choose this value as an optimal width of a Gaussian kernel. 

\section{Conclusion}
In conclusion, we have proposed a new approach for analyzing GS in a unified framework of a kernel method.
We have tested the proposed approach by applying it to several examples exhibiting GS, and demonstrated that the canonical correlation coefficient of the Kernel CCA is a suitable  index for the characterization of GS.
In addition, it has been shown that nonstationary changes of the coupling  are detected from the time series by the difference between canonical variates of the Kernel CCA.
It has been also discussed  how the parameter of the kernel function can be suitably chosen from data by the procedure of cross-validation. 
Our experiments show that a method based on  CV gives promising results in optimizing the parameter $\sigma$. 
The cross-validation procedure is also useful to circumvent spurious detection of GS by overfitting. 

The approach based on the Kernel CCA provides not only an index for measuring nonlinear interdependence.  
It also provides global nonlinear coordinates, and these coordinates allow a representation of the interaction between the dynamical systems under investigation. 
Note that the linear CCA also provides global coordinates, but it cannot discriminate between linear and nonlinear relation. 
In this respect,  it goes beyond conventional methods of analyzing GS.  
Our attempts open a new possibility of the kernel methods for analyzing complex dynamics observed in nonlinear systems.  
 
\ack
We thank S.~Akaho  
for stimulating discussions on the kernel methods,  
and K.~Fukumizu for useful comments and informing us his results prior to publication.
 We also thank Y.~Hirata for useful suggestions on surrogate data analysis.
 

\end{document}